\documentclass[
aps,%
12pt,%
final,%
notitlepage,%
oneside,%
onecolumn,%
nobibnotes,%
nofootinbib,%
superscriptaddress,%
noshowpacs,%
centertags]%
{revtex4}

\usepackage{epsfig}
\usepackage[russian]{babel}

\begin{document}

\begin{center}

\textbf{ A.N. Sazonov\\Optical Multicolor Observations of the  SS
433=V 1343 Aql Microquasar}\\

 2009  A.N. Sazonov

Sternberg Astronomical Institute, Universitetskii pr. 13, Moscow, 119992 Russia

\end{center}

~ We report BVR photometry of the V1343 Aql= SS 433 microquasar at
different phases of the 13-day orbital cycle for the 1986-1990
observing seasons. The data include five complete cycles of the
163$^{d}$ precession period of the system. We obtain mean light
curves and color-color diagrams with the orbital period for all
intervals of precession phases. The optical component of the close
binary system (CBS) fills its critical Roche lobe and loses mass
on the thermal relaxation time scale. Gaseous flow show up
actively in the system and activity manifestations differ
substantially at different precession phases.

~ The collimated relativistic jets perpendicular to the plane of
the disk appear to be associated with supercritical accretion onto
the compact relativistic object in the massive CBS, which shows up
in the shapes of the light curves at different orbital and
precession phases. An analysis of color indices confirmed the
earlier discovered peculiarities of the system
(~\cite{Shakhovskoi1996}): (1) A  "disk corona" around the compact
object. (2) Phase shifts between orbital light curves and
different heights of light maxima for different passbands and at
all phases of the 163$^{d}$ precession period.

~ The BVR light curves of the CBS V1343 Aql= SS 433 are
qualitatively similar, the depth of the primary minimum (the
eclipse of the accretion  disk of the system by it ``normal''{} -
star component) during the observing seasons considered was of
about 0$^{m}$.65$\div $0$^{m}$.75, 0$^{m}$.50$\div $0$^{m}$.60,
and 0$^{m}$.35$\div $0$^{m}$.45 in the B, V, and R passbands,
respectively. The light curves vary with the precession phase in
what must be a manifestation of the activity of gaseous flows of
the CBS and their active interaction with the ``floating'' {}
accretion disk whose position in space is related with the
orientation of relativistic jets.

~ The phase shifts of orbital light curves and different heights
of light maxima confirm the asymmetry of the brightness
distribution in the accretion disk and the fact that the
downstream side of the AD (with respect to orbital motion) is
brighter than the upstream side.

~ Of considerable interest and value are multicolor photometric
observations of SS 433~\cite{Shakhovskoi1996},
~\cite{Sazonov1988}), because most of the extensive V-band, and
much scarcer B and R-band observations of this unique object were
made in only, mostly the V-band, filter~\cite{Cherepashchuk1981a};
~\cite{Kemp1986}; ~\cite{Cherepashchuk1991};
~\cite{Irsmambetova1997};

~The color variations of the star with the orbital and precession
periods at  Min I and Min II have been confirmed conclusively
during the observing seasons considered .

~ \cite{Shakhovskoi1996} obtained a number of original results
obtained for the first time with the AZT-11 telescope equipped
with a five-color UBVRI photometer of the Helsinki
University~\cite{Piirola1973}; ~\cite{Shakhovskoi1996};
~\cite{Korhonen1983} with a large aperture.

~ The device can be used to perform synchronous polarimetric and
photometric observations in five bands close to the standard bands
of the Johnson's UBVRI system, which is very important for a
quantitative and qualitative comparison with other observations
made in the same photometric systems.

~ During the 1986-1990 observing period a series of five to seven
outbursts occurred in each observing season (especially in 1988)
with the V-band amplitudes ranging from $\sim $0$^{m}$.30 to $\sim
$0$^{m}$.50. Optical outbursts of  SS 433 had almost uniform
distribution of precession phases during these observing seasons.

~ Observations were performed on time scales of fast variability
with exposures ranging from $90^s$ to $180^s$ on time intervals
from 1 to 2.5 hours during the night.

~ The small amplitude of B-V color index variations shows up
conspicuously in the color-index diagrams. It is evidently due to
the relatively small B-V color differences between the thermal
continuum emitted by the bright optical star and different parts
of the accretion disk of the system (~\cite{Shakhovskoi1996}),
because above 20000~K (B-V) colors are practically independent of
temperature.

~ Note that the behavior of periodic light variations of SS 433
during these observing seasons qualitatively agrees with the data
by other authors for the period from 1978 through 1990, suggesting
a certain stability of regular light variations of the system with
the periods of 163$^{d}$.34 (the precession period, the amplitude
$\Delta $V$ \cong $0$^{m}$.75$\div $0$^{m}$.85) and 13$^{d}$.086
(the orbital period, the corresponding amplitude is $\Delta $V$
\cong $0$^{m}$.75$\div $0$^{m}$. 85).

~ We also studied the nutation light curve of the object based on
observations made in 1986-1990, i.e., the light curve of
variations that remain after the subtraction of precession and
orbital variations and folded with the nutation period of 6.28~d.
The computed ephemeris for the maximum light of nutation
variations is (~\cite{Goranskii1998a}):

$$
 Max = JD2450000^d.94 + 6^d.2877 E
$$

~ \cite{Kopylov1987} pointed out that one of the most salient
features of the behavior of relativistic jets as a function of the
phase of the nutation cycle is the sharp increase of the
equivalent width W$_{\lambda} $ of relativistic lines at the
bending points of the radial-velocity curves and that this this
effect is especially conspicuous at phases $\sim  \quad \psi
$=0.00, where jets are most inclined to the line of sight of a
terrestrial observer. Moreover, at these phases variation of the
angle between the line of sight and direction of the jet is little
affected by the precession motion of relativistic jets and the
angle considered varies mostly due to nutational oscillation of
the jets. Unpredictable enhancements of line components are
sometimes observed at other nutational phases (e.g., the
H$_{\alpha} ^{ -} line)$. There is a viewpoint that these features
are due to the nonstationary nature of the activity of SS 433 and
strong variations of the mass-loss rates in relativistic jets.
However, on the other hand, there may be other singular points
besides the bending points on the radial-velocity curves, e.g.,
\textbf{\textit{self-crossing points of the jets of the gaseous
stream.}}

~ \cite{Sazonov1988} also pointed out the above property of the
trajectories of the particles of the gaseous stream in a CBS ---
the intersection (even in the celestial-mechanics approximation).
The fact that ballistic trajectories of stream particles intersect
may prove to be one of the factors conducive to the formation of
irregularities in the gaseous jet, which result in scintillations
of the brightness of the ``hot spot'' that forms where the jet
encounters the disk-like envelope of the relativistic object of
the CBS, and also in the case of self-crossing of the jet
trajectory including the portions of the collimated-jet trajectory
near T3 phase (more precisely, at phases $\psi=0.03-0.15$).

~ \cite{Bisikalo1994}; ~\cite{Bisikalo1996} and ~\cite{Lubow1993}
in his review point out various gaseous-flow instabilities near
the compact object in the CBS. The above authors also point out
that observed manifestations of gas flows include quasi-periodic
oscillations of accretion rate М and the rate of change of the
angular momentum, J, of the matter located in the vicinity of the
secondary component of the CBS. The computed density, velocity,
and temperature fields were used to determine the emission
profiles of the  H$_{\beta} $ and establish that line wings form
in the region near the accretor and that their broadening is
determined by high velocities of the gas flow in the accreting
matter and in the disk. \textbf{\textit{Gas velocities near the
accretor are more than 10 times higher than than the average gas
velocity in the system}} ~\cite{Bisikalo1996}.

~ It was also established earlier that gas in the jet concentrates
 in individual clouds~\cite{Shklovskii1981};~\cite{Murdin1980};
~\cite{Grandi1982}) a pattern, which is consistent with the
scenario proposed to explain the observed data points with
somewhat higher-than-average amplitude at precession phases $\psi
$=0.02-0.16 (and at different orbital phases $\phi )$ in SS 433.

\end{document}